\newif\ifarxiv
\arxivtrue
\ifarxiv%(
\RequirePackage{snapshot}
\documentclass[
    reprint,
    superscriptaddress,
    amsmath,amssymb,
    aps,
    pra,
]{revtex4-2}
\renewcommand{\doi}[1]{\href{http://doi.org/#1}{doi:#1}}
\usepackage{pgffor}
\else%)(
\documentclass{WileyChemistry-template}
\fi%)
\newif\ifshowlabels
\showlabelsfalse
\newif\iffigdraft
\figdraftfalse
\newif\ifgentoc
\newif\ifeditmode
\editmodefalse
\gentocfalse
\newif\ifshowparagraphinline
\showparagraphinlinefalse
\ifeditmode\else
\showparagraphinlinefalse
\fi
\ifarxiv%(
\usepackage[sectionbib]{bibunits}
\defaultbibliographystyle{new}
\defaultbibliography{references}
\else%)(
\fi%)
\usepackage{layouts}% for printinunitsof and plrntlen
\usepackage[acronyms]{glossaries}
\glsdisablehyper
\usepackage[procnames]{listings}
\usepackage{amsfonts}
\usepackage{blindtext}
\let\oldblindtext\blindtext
\renewcommand{\blindtext}{\textcolor{gray}{\oldblindtext}}
\usepackage{cmap}% so that I can copy and paste -- in word, use <C-H>, then replace ^p with space -- also allows ligatures to be searched
\usepackage{datetime}
\usepackage{environ}
\usepackage{etoolbox}
\usepackage[utf8]{inputenc}
\DeclareUnicodeCharacter{03BD}{\ensuremath{\nu}}%$\nu$}
\DeclareUnicodeCharacter{2212}{--}%$\nu$}
\DeclareUnicodeCharacter{03BC}{\textmu}%$\mu$}
\DeclareUnicodeCharacter{03B2}{\ensuremath{\beta}}%$\nu$}
\DeclareUnicodeCharacter{03B1}{\ensuremath{\alpha}}%$\nu$}
\DeclareUnicodeCharacter{00B0}{\ensuremath{^\circ}}%$\nu$}
\DeclareUnicodeCharacter{00B1}{\ensuremath{\pm}}%$\nu$}
\DeclareUnicodeCharacter{2192}{\ensuremath{\rightarrow}}%$\nu$}
\DeclareUnicodeCharacter{03C3}{\ensuremath{\sigma}}%$\sigma$}
\DeclareUnicodeCharacter{03C4}{\ensuremath{\tau}}%$\sigma$}
\DeclareUnicodeCharacter{2009}{ }%$\nu$}
\DeclareUnicodeCharacter{2080}{\ensuremath{_0}}
\DeclareUnicodeCharacter{2082}{\ensuremath{_2}}% ₂
\DeclareUnicodeCharacter{2084}{\ensuremath{_4}}
\DeclareUnicodeCharacter{1F449}{\ensuremath{\rightarrow}} % pointing finger
\DeclareUnicodeCharacter{1F448}{\ensuremath{\leftarrow}} % pointing finger
\iffigdraft
    \usepackage[draft]{graphicx}
\else
    \usepackage{graphicx}
\fi
\usepackage{hyphenat}
\usepackage{ifxetex}
\usepackage{mathrsfs}% script fonts mathscr
\usepackage{calc}

\newcommand{\ie}{\textit{i.e.\@}\xspace}
\newcommand{\vs}{\textit{vs.\@}\xspace}
\newcommand{\eg}{\textit{e.g.\@}\xspace}
\newcommand{\etc}{\textit{etc.\@}\xspace}

\newcommand{\relxvtunits}{\ensuremath{\text{M}^{-1}\text{s}^{-1}}}

\usepackage{hyperref}
\hypersetup{colorlinks=true, pdfstartview=FitV, linkcolor=linkcolor, citecolor=dbluecolor, urlcolor=dgreencolor, bookmarksdepth=subparagraph}% hyperref for pdf -- no idea what half these are, but the bookmarksdepth is to include down to subparagraph
\usepackage{bookmark}
\usepackage{sidecap}
\usepackage{soul}
\usepackage{textcomp}% among other things, gives \textmu and \textperthousand
\usepackage{url}
\usepackage{xcolor}
\usepackage{xspace}
\usepackage{multirow}
\usepackage{array}% enables >{\displaystyle}, etc., as well as *2, etc.
\newcolumntype{R}{>{\vspace{2ex}\displaystyle}{r}}
\newcolumntype{L}{>{\displaystyle}{l}}
\definecolor{SUgrey}{HTML}{6F777D}
\definecolor{SUorange}{HTML}{D44500}
\definecolor{dbluecolor}{rgb}{.01,.02,0.29}
\definecolor{dgraycolor}{rgb}{0.50,0.50,0.50}
\definecolor{dgreencolor}{rgb}{0.0,0.4,0}
\definecolor{linkcolor}{cmyk}{0,0.7,0.5,0.5}
\usepackage{accsupp}    
\newcommand{\noncopynumber}[1]{%
    \BeginAccSupp{method=escape,ActualText={}}%
    #1%
    \EndAccSupp{}%
}
\ifarxiv%(
\usepackage[compact]{titlesec}
\fi%)
\ifshowparagraphinline%(
\titleformat{\paragraph}[runin]{\color{gray}\normalfont\bfseries\footnotesize}{}{3pt}{\hspace{0.75em}\ul{\footnotesize\thesubsection\theparagraph)\;}}[:]%\titleformat{command}[shape (eg. runin)]{format}{label}{sep}{before}[after]
\else%)(
\renewcommand{\paragraph}[1]{\par\phantomsection\addcontentsline{toc}{paragraph}{#1}}
\fi%)
\makeatletter
\@ifundefined{linelabel}{%
\newcommand{\linelabel}[1]{}}{}
\makeatother
\makeglossaries
\renewcommand{\glossarysection}[2][]{}
\newacronym{odnp}{ODNP}{Overhauser Dynamic Nuclear Polarization}
\newacronym{adrosys}{ADROSYS}{Automated Deuterium Relaxation-Ordered SpectroscopY in Solution}
\newacronym{snr}{SNR}{signal-to-noise ratio}
\newacronym{nmr}{NMR}{Nuclear Magnetic Resonance}
\newacronym{esr}{ESR}{Electron Spin Resonance}
\newacronym{mps}{MPS}{Microwave Power Source}
\newacronym{ilt}{ILT}{Inverse Laplace Transform}
\newacronym{pre}{PRE}{Paramagnetic Relaxation Enhancement}
\newacronym{rosy}{ROSY}{Relaxation-Ordered SpectroscopY}
\newacronym{dss}{DSS}{Dynamic Stokes Shift}
\newacronym{md}{MD}{molecular dynamics}
\newacronym{ir}{IR}{infrared}
\newacronym{tms}{TMS}{tetramethylsilane}
\newacronym[plural=SPs,firstplural=electron spin probes (SPs)]{sp}{SP}{electron spin probe}
\newacronym[plural=RMs,firstplural=Reverse Micelles (RMs)]{rm}{RM}{reverse micelle}
\newacronym{aot}{AOT}{Aerosol-OT}
\newacronym{qens}{QENS}{Quasi-Elastic Neutron Scattering}
\renewcommand{\thesection}{\Roman{section}}
\renewcommand{\thesubsection}{\thesection.\arabic{subsection}}

\makeatletter
\renewcommand{\p@subsection}{}
\renewcommand{\p@subsubsection}{}
\makeatother
\newcounter{subfigure}[figure]
\newcounter{subfigurenonumber}
\newcounter{tempfigure}
\renewcommand\thesubfigurenonumber{(\alph{subfigurenonumber})}% wrapping with textbf causes problems
\newcommand{\subfig}[2]{%
    \setcounter{tempfigure}{\value{figure}}%
    \addtocounter{tempfigure}{1}%
    \refstepcounter{subfigure}%
    \setcounter{subfigurenonumber}{\value{subfigure}}%
    \expandafter\edef\csname ref#2\endcsname{\thesubfigurenonumber}
    \label{#1}%
    }
\ifarxiv%(
\newcommand{\titleblock}{% {{{ title info
\newcommand\SUaffil{\affiliation{Department of Chemistry, Syracuse University, Syracuse, NY 13210, USA}}
\author{Alec A. Beaton}
\author{John M. Franck}
\SUaffil
\email{jmfranck@syr.edu}
\title{Direct Observation of the Translational Immobilization of Water Molecules Under Nanoscale Confinement}
\date{\today}
}% }}}
\else%)(
\newcommand{\authorList}{
    {Alec A. Beaton},
    {John M. Franck$^*$}
}
\newcommand\SUaffil{Department of Chemistry, Syracuse University, Syracuse, NY 13210, USA}
\newcommand\myemail{jmfranck@syr.edu}
\fi%)

\newcommand{\wileyAuthors}[1]{
    \foreach \a [count=\i from 1] in #1 {%
        \ifnum \i=1%
            \a\textsuperscript{[a]}%
        \else%
            , \a\textsuperscript{[a]}%
        \fi%
    }
}

\newcommand{\mytitle}{Direct Observation of the Translational Immobilization of Water Molecules Under Nanoscale Confinement}

\ifarxiv\else%(
    \title{\mytitle}
    \author{
        \begin{minipage}{\textwidth}
            \centering
            \wileyAuthors{\authorList}
        \end{minipage}
    }
    \newcommand{\affiliation}{
    \begin{itemize}

    \item[{[a]}] \foreach \a [count=\i from 1] in \authorList {\ifnum \i=1 \a\else, \a\fi}
        \SUaffil\\
    E-mail: \myemail
    \end{itemize}
    }
\fi%)
\newif\ifpoormancref
\poormancreffalse
\ifpoormancref
\usepackage[poorman]{cleveref}
\else
\usepackage{cleveref}%for cref
\fi
\crefname{equation}{Eq.}{Eqs.}
\crefname{table}{Table}{Tables}
\crefname{figure}{Fig.}{Figs.}
\crefname{section}{Sec.}{Sec.}
\crefname{subfigure}{Fig.}{Figs.}
\crefname{lstlisting}{listing}{listings}
\Crefname{lstlisting}{Listing}{Listings}

\usepackage{xr}
\ifshowlabels
\usepackage{refcheck}
\makeatletter
\newcommand{\refcheckize}[1]{%
  \expandafter\let\csname @@\string#1\endcsname#1%
  \expandafter\DeclareRobustCommand\csname relax\string#1\endcsname[1]{%
    \csname @@\string#1\endcsname{##1}\wrtusdrf{##1}}%
  \expandafter\let\expandafter#1\csname relax\string#1\endcsname
}
\def\@setmarginlbl{%
    \if@show@ref
        \if@labelled
            \set@fbox@par
            \if@unsdlbl
                \makebox[0pt][l]{\zero@height{$\,$\rotatebox{90}{\scalebox{0.6}{\mark@size
                {\bfseries\upshape?}\underline{\last@lbl}{k\bfseries\upshape?}}}}}%
            \else
                \makebox[0pt][l]{\zero@height{$\,$\rotatebox{90}{\scalebox{0.6}{\fbox{{\mark@size\last@lbl}}}}}}%
            \fi
        \else
            \if@show@unl@bld
                \makebox[0pt][l]{\zero@height{$\,$\rotatebox{90}{\scalebox{0.6}{\unl@bld@mark}}}}%
            \fi\fi
        \fi
        \global\@labelledfalse
    }
\def\@setnmmarginlbl{%
    \if@show@ref
        \set@fbox@par
        \if@unsdlbl
            \hbox to \textwidth{\makebox[0pt][r]{\rotatebox{90}{\scalebox{0.6}{\mark@size{\bfseries
                            \upshape?}$\langle$\last@lbl$\rangle${\bfseries
            \upshape?}}}$\,$}\hfill}%
        \else
            \hbox to \textwidth{\makebox[0pt][r]{\rotatebox{90}{\scalebox{0.6}{\mark@size$\langle$%
            \last@lbl$\rangle$}}$\,$}\hfill}%
        \fi
    \fi
    \global\@labelledfalse
}
\def\@bibitem@proceed@#1{%
    \@ifundefined{cit@#1}{\@warning@rc@{Unused bibitem `#1'}%
        \if@show@cite
            \gdef\@biblabel{\makebox[0pt][r]{\zero@height{\rotatebox{90}{\scalebox{0.6}{{\mark@size{\bfseries\upshape?}}%
                \underline{\@verbatim@{#1}}{\mark@size{\bfseries\upshape?}}}}$\,$}}%
            \@@@biblabel@@}%
        \fi
    }{%
        \if@show@cite
            \set@fbox@par
            \gdef\@biblabel{\makebox[0pt][r]{\zero@height{\rotatebox{90}{\scalebox{0.6}{\fbox{TESTTESTTEST\@verbatim@{#1}}}}$\,$}}\@@@biblabel@@}%
        \fi
}}%
\makeatother
\refcheckize{\cref}
\refcheckize{\Cref}
\fi
\begin{document}
\newcommand{\figPREpieces}{\begin{figure*}[tbp]
    % from enc slides
    \centering
    \includegraphics[width=\linewidth]{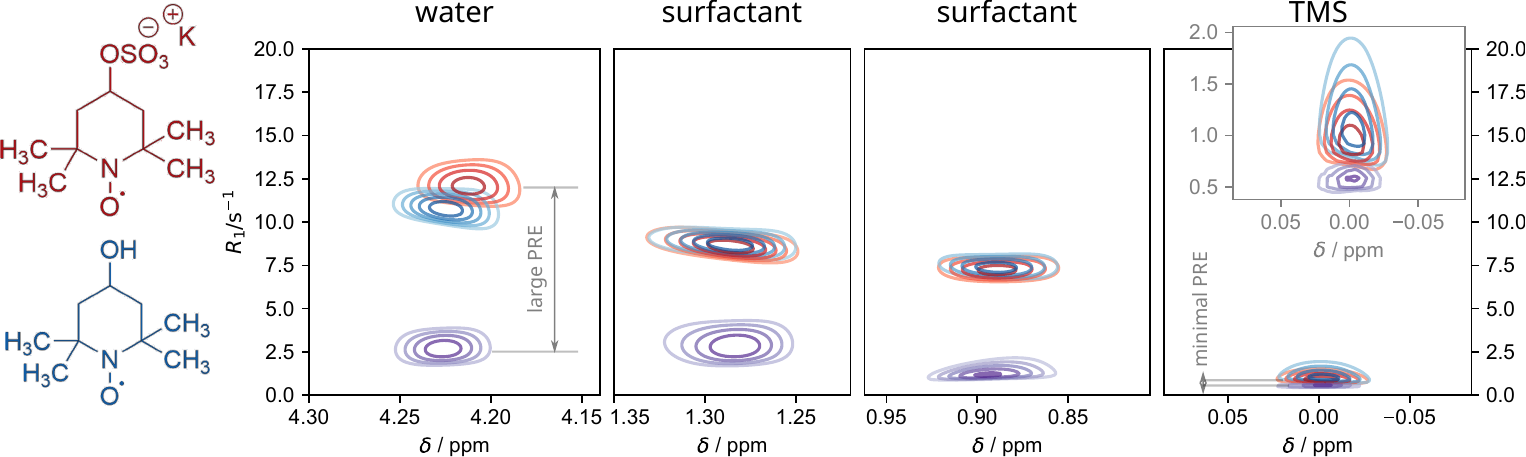}
    \caption{
    The water, surfactant (2 strips), and \gls{tms}
    regions of the \gls{pre}-\gls{rosy} spectra for \gls{rm} samples that contain water with no \gls{sp}
    (purple), \vs 70~mM TEMPOL (blue) or 70~mM TEMPO-SO$_{4}$ (red).
    The displacement of the $R_1=1/T_1$ in solutions containing
    spin label (red and blue) relative to the control
    (purple) is the \gls{pre} effect, and indicates
    the extent to which
    the species of interest approaches the
    nitroxide group of the spin label.
    The inset box (top right, grey border) zooms in on the \gls{tms} signal.
}
    \label{fig:PREpieces}
\end{figure*}}

\newcommand{\figPRE}{\begin{figure*}[tbp]
    % from enc slides
    \centering
    \begin{tabular}{cccc}
    \includegraphics[width=0.75\linewidth]{figures/PRE_ILT_all.png}
    \end{tabular}
    \caption{
    The T$_{1}$ results performed for \gls{pre} analysis, with
    for \gls{rm} samples prepared with just water
    (purple), 70~mM TEMPOL (blue), and 70~mM TEMPO-SO$_{4}$ (red).}
    \label{fig:PREfull}
\end{figure*}}

\newcommand{\figkSigma}{\begin{figure}[tbp]
    % from enc slides
    \centering
    \subfig{fig:Ep}{Ep}
    \subfig{fig:ksigma}{kSigma}
    \begin{tabular}{cc}
        \refEp &
        \hspace*{-2em}\raisebox{-\height}{\includegraphics[width=0.9\linewidth]{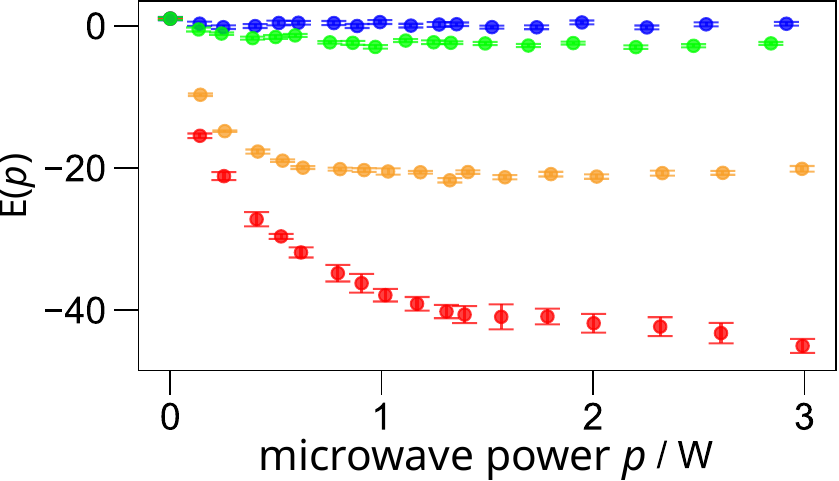}}
        \\
        \refkSigma &
        \hspace*{-2em}\raisebox{-\height}{\includegraphics[width=0.9\linewidth]{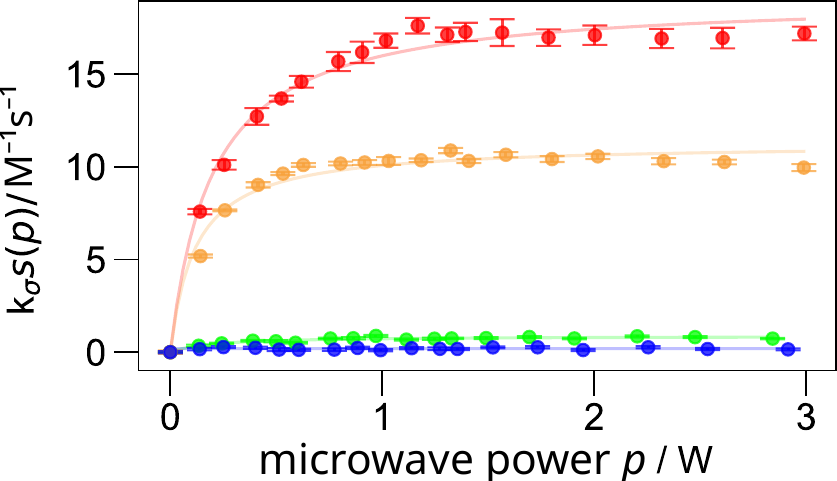}}
    \end{tabular}
    \caption{
        \refEp\ 
        \gls{nmr} signal enhancement,
        as a function of \gls{esr}-resonant microwave power,
        for the \gls{aot}/CCl$_{4}$ \gls{rm}s for $w_{0}$ = 1 (blue), 5 (green), 7.4 (orange), and 10 (red). Each has been corrected for the corresponding enhancement of the \gls{aot}.
        Dividing out the \gls{nmr} $T_1$ relaxation times and
        several constants
        (\cref{eq:polarization_subst_fundamental0})
        yields the product of the
        \gls{esr} saturation factor
        and the desired cross-relaxation rate $k_\sigma$,
        which is the asymptotic limit of the fit curve in
        \refkSigma,
        and which directly
        indicates substantial variation of the translational motion
        with $w_0$.
    }
\end{figure}}

\newcommand{\figRMESR}{\begin{figure}[tbp]
    % from enc slides
    \centering
    \includegraphics[width=\linewidth]{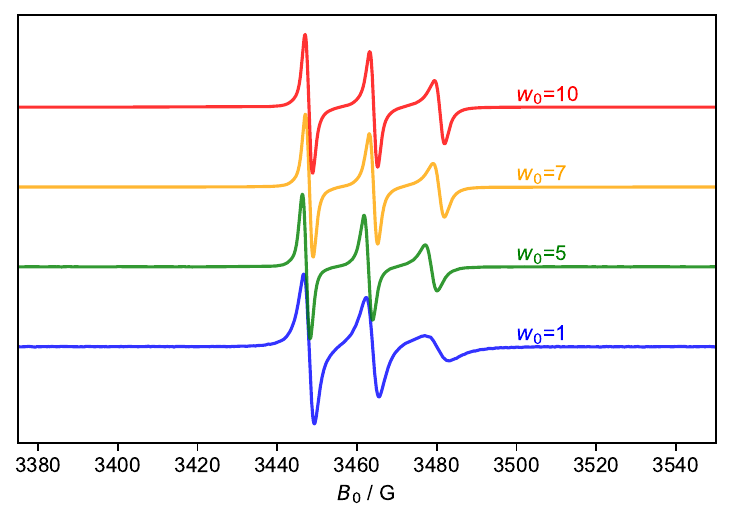}
    \caption{
        Normalized \gls{esr} spectra acquired for the \gls{rm}
        samples containing TEMPO-SO$_{4}$ in
        H$_{2}$O for w$_{0}$ = 1 (blue), 5
        (orange), 7.6 (green), and 10 (red).
    }
    \label{fig:rm_esr}
\end{figure}}
\newcommand{\tabRMESR}{\begin{table}
    \centering
\begin{tabular}{cc}
$w_0$ & $\tau_c/\text{ns}$\\
\hline\hline
      1 & 1.7\\
5 & 0.51\\
7 & 0.44\\
10 & 0.35\\ \hline
\end{tabular}
\caption{\gls{esr} correlation times corresponding to the spectra in \cref{fig:rm_esr}.}
\label{tab:RMESR}
\end{table}}

\newcommand{\figCarbonTetSL}{\begin{figure}[tbp]
    % (all are auto apodized, saved as respective
    % processed files, then plotted, saved as pdf,
    % opened in inkscape and labels manually
    % edited)
    % (reference enc-2022 folder in lab notebook
    % if unsure)   
    \centering
    \subfig{fig:justCtet}{justCtet}
    \subfig{fig:CtetwithSL}{CtetwithSL}
    \begin{tabular}{cc}
    \refCtetwithSL &
    \includegraphics[width=0.9\linewidth]{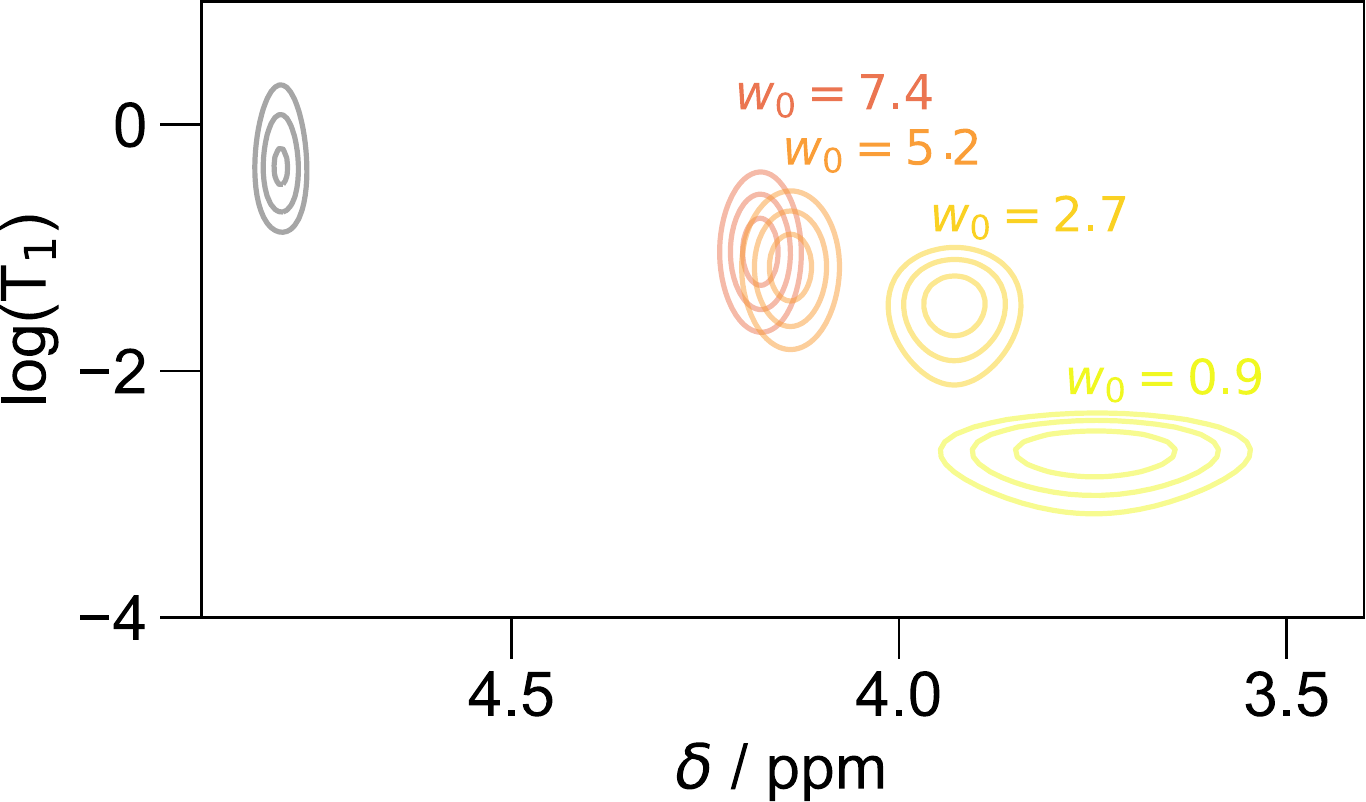}
    \\
    \refjustCtet &
    \includegraphics[width=0.9\linewidth]{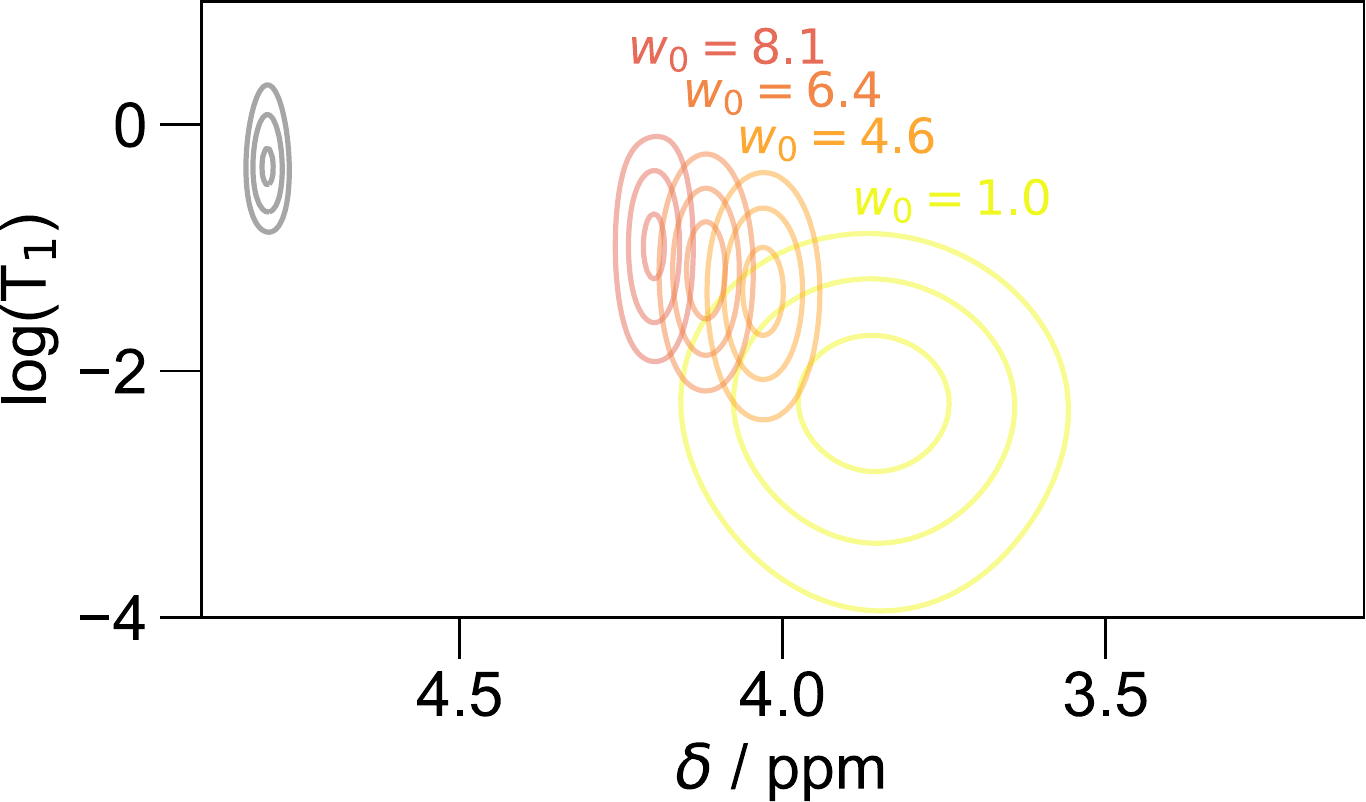}
    \end{tabular}
    \caption{ILT plot showing the T$_{1}$ of
        D$_{2}$O for different \gls{aot} \glspl{rm} prepared in
        CCl$_{4}$ \refjustCtet\ and in
        CCl$_{4}$ with added spin probe
        TEMPO-SO$_{4}$ \refCtetwithSL\ 
        for water loadings ranging
        from 1 to 8.
    }
    \label{fig:CCl4T1SL}
\end{figure}}

\newcommand{\figCompareEnh}{\begin{figure}[tbp]
    % from enc slides
    \centering
    \includegraphics[width=\linewidth]{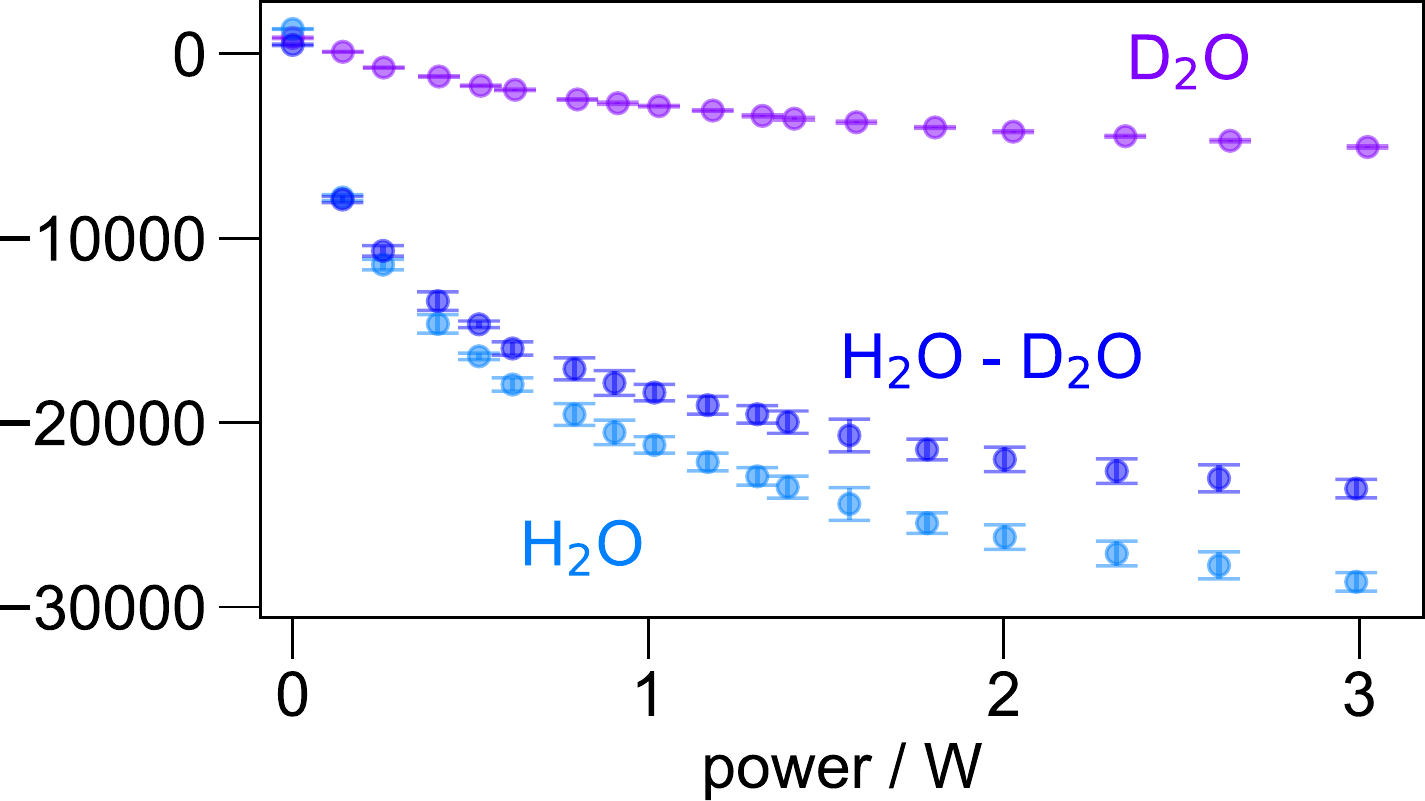}
    \caption{
    The unnormalized signal intensity curves for $w_{0}$ = 10 \gls{aot}/CCl$_{4}$
    \gls{rm}s, showing the difference between a sample prepared with H$_{2}$O vs
    D$_{2}$O.
    The D$_{2}$O data contains only enhanced
    signal from the surfactant.
    The units of the $y$-axis are arbitrary
    units for the integrated \gls{nmr} signal intensity.
    The subtraction is performed on the unnormalized data and also
    shown;
    the subtracted data is subsequently
    normalized by the value at $p=0$ to
    generate the appropriate $E(p)$ curve.}
    \label{fig:CompareEnh}
\end{figure}}

\newcommand{\figRadiusPlot}{\begin{figure}[tbp]
    % py plot_ksigma_distance.py
    \centering
    \includegraphics[width=\linewidth]{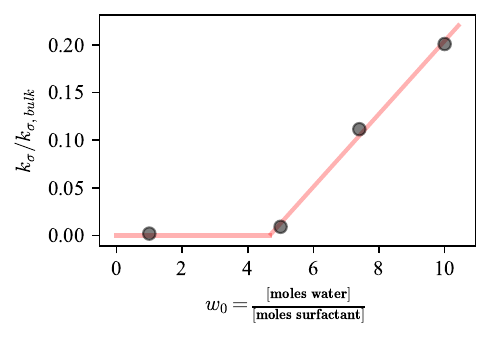}
    \caption{The variation of translational mobility,
        as quantified by the \gls{odnp} cross-relaxivity
        (normalized again the cross-relaxivity of a small
        spin label dissolved in bulk water $k_{\sigma,bulk}$)
        as a function of the degree of confinement--given here by the
        water loading $w_0$.
        Note that the diameter of the water pool can be approximated
        by $w_0\times 0.35\;\text{nm}$.
    }
    \label{fig:RadPlot}
\end{figure}}
\newcommand{\tabKsig}{%
\begin{table}
    \begin{center}
        \resizebox{\linewidth}{!}{%
                \begin{tabular}{c|ccc}
                    \textbf{$w_{0}$}
                    &
                    \quad
                    \textbf{$k_{\sigma}/M^{-1}s^{-1}$}
                    \quad
                    &
                    \quad
                    \textbf{$k_{\sigma,bulk}/k_\sigma$}
                    \quad
                    &
                    \quad
                    \textbf{$k_{low}/k_{low,bulk}$}
                    \quad
                    \\ 
                    \hline
                    \hline
                    1       & 0.19 & 320& $>0.81$\\
                    5       & 0.87 & 68 & $>0.84$\\
                    7.4     & 10.6 & 8.4& $>0.99$\\
                    10      & 19.1 & 5.0& $0.99$ \\
                \end{tabular}
        }
        \caption{Values for $k_{\sigma}$ and $k_{low}$ determined for different \gls{rm} samples,
            with $k_{\sigma,bulk}=95.4\;\relxvtunits$ and $k_{low,bulk}=366\;\relxvtunits$ \cite{FranckMethEnz2018}.
    }
    \label{tab:table_odnp}
    \end{center}
\end{table}%
}

\ifarxiv\else%(
\twocolumn[\vspace{-1.5cm}\maketitle\vspace{-1cm}
	\textit{\dedication}\vspace{0.4cm}]
\small{\begin{shaded}
        \noindent\Gls{odnp}
    detects 
    an experimental measurable associated directly with translational
    motion of water at the nanoscale,
    a quantity that few other methods can detect.
This study offers a unique insight into the
    translational diffusion of water inside \glspl{rm}.
It finds
    that
    simply adjusting the
    ``water loading'' ($w_0$, \textit{i.e.} the mole ratio of
    surfactant to water)
    to tune the size of the \glspl{rm}
    achieves a near-continuous tuning of the
    translational diffusion of water.
Furthermore,
    (1)
    the dynamics of water molecules in the core of relatively large \glspl{rm}
    ($w_0=10$, diameter of water nanopool $\approx 3.5\;\text{nm}$)
    diffuse only as fast
    as those on the surface of a lipid bilayer
    and (2)
    surprisingly, translational diffusion slows to
    a near-stop for \glspl{rm} that are small,
    but that still contain hundreds of water molecules
    in their core.
Extrapolation to larger sized water pools implies
    that in order to recover
    bulk-like translational dynamics,
    tens of thousands of water molecules
    are required.
The data from the small \glspl{rm}
    also represent a breakthrough
    as the first example where
    a spin probe that is completely exposed to
    water (as opposed to buried inside a macromolecule)
    observes dramatic slowing of the translational diffusion.

	\end{shaded}
}
\begin{figure} [!b]
\begin{minipage}[t]{\columnwidth}{\rule{\columnwidth}{1pt}\footnotesize{\textsf{\affiliation}}}\end{minipage}
\end{figure}
\fi%)
% {{{ latexdiff directives?
%REMOVE \let\oldalignstart=\align
%REMOVE \let\oldalignend=\endalign
%REMOVE \let\oldequationstart=\equation
%REMOVE \let\oldequationend=\endequation
%REMOVE \let\oldeqnarraystart=\eqnarray
%REMOVE \let\oldeqnarrayend=\endeqnarray
%REMOVE \let\oldintertext=\intertext
%REMOVE \renewenvironment{align}{\endlinenumbers\oldalignstart}{\oldalignend\linenumbers}
%REMOVE \renewenvironment{equation}{\endlinenumbers\oldequationstart}{\oldequationend\linenumbers}
%REMOVE \renewenvironment{eqnarray}{\endlinenumbers\oldeqnarraystart}{\oldeqnarrayend\linenumbers}
%REMOVE \renewcommand{\intertext}[1]{\oldintertext{\begin{linenumbers}#1\end{linenumbers}}}
%REMOVE \diffexpl
%REMOVE 
%REMOVE \begin{linenumbers}
% }}}
\newlength\myfigwidth
\setlength{\myfigwidth}{3.5in}
\ifarxiv%(
\titleblock

\begin{abstract}
\Gls{odnp}
    detects 
    an experimental measurable associated directly with translational
    motion of water at the nanoscale,
    a quantity that few other methods can detect.
This study offers a unique insight into the
    translational diffusion of water inside \glspl{rm}.
It finds
    that
    simply adjusting the
    ``water loading'' ($w_0$, \textit{i.e.} the mole ratio of
    surfactant to water)
    to tune the size of the \glspl{rm}
    achieves a near-continuous tuning of the
    translational diffusion of water.
Furthermore,
    (1)
    the dynamics of water molecules in the core of relatively large \glspl{rm}
    ($w_0=10$, diameter of water nanopool $\approx 3.5\;\text{nm}$)
    diffuse only as fast
    as those on the surface of a lipid bilayer
    and (2)
    surprisingly, translational diffusion slows to
    a near-stop for \glspl{rm} that are small,
    but that still contain hundreds of water molecules
    in their core.
Extrapolation to larger sized water pools implies
    that in order to recover
    bulk-like translational dynamics,
    tens of thousands of water molecules
    are required.
The data from the small \glspl{rm}
    also represent a breakthrough
    as the first example where
    a spin probe that is completely exposed to
    water (as opposed to buried inside a macromolecule)
    observes dramatic slowing of the translational diffusion.

\end{abstract}
\begin{bibunit}
%\keywords{ESR, ODNP, NMR, cross-relaxation}
\maketitle
\glsresetall
\fi%)
\section*{Introduction}
\label{sec:ODNPRMintro}
Polymer pores,
    pockets of proteins,
    and solid-phase meso- and nano-porous systems
    can all confine water,
    dramatically altering its
    properties~\cite{Laage2017,Zhong_cpl,Fardis2022SelConWat,Persson2018HowProMod,KangHu2023DynEleFie}
    and, therefore, how it contributes to and drives
    chemistry~\cite{Munoz2021ConAquChe,Qian2020NanWatTre,Chong2017,ChoSitSitBac2021,Bellissent2016WatDetStr}.
\Glspl{rm} are an important model system for
    understanding confined water: one of only a few
    systems where the size of enclosure can be changed
    almost continuously without changing the chemical identity of the
    enclosing
    structure~\cite{Cerveny2016,biswasAnomalous2018,tanDynamics2005,Fardis2022SelConWat,Fayer2010AnaWatCon,sanfordSweet2018}.
Previous measurements of \gls{rm} dynamics include
    infrared spectroscopy~\cite{pileticTesting2006}
    \gls{qens} coupled with \gls{md}
    simulations~\cite{harphamWater2004},
    \gls{dss}~\cite{Sarkar1996SolDynCou,Singha2014SelProRev,Faeder2005SolDynRev},
    pump-probe spectroscopy of
    \gls{ir}-active probe molecules~\cite{Lee2014InfPumPro,Roy2017},
    deuterium
    or $^{17}$O
    \gls{nmr}~\cite{carlstroemShape1989,carlstroemWater1988a,quistWater1988},
    and dielectric spectroscopy~\cite{DAngelo1996}.
All tend to indicate reduced mobility of the confined
    water,
    but it can be difficult to
    experimentally isolate the 
    translational diffusion of the confined water;
    furthermore,
    especially when the motion is slow,
    distinguishing the motion of the water from that of
    the \gls{rm} can prove challenging.
Isolating the translational motion of
    confined water is crucial,
    since it is believed to be both decoupled from
    other motions~\cite{PanTan2020DecTraRot},
    as well as essential,
    \eg, to the association of macromolecules
    and the domain motion of
    proteins~\cite{Schiro2015,SchiroRolHydWat2019},
    or to explaining important transport
    properties in energy-relevant
    materials~\cite{Zhang2020RelGeoNan}.

\Gls{odnp} reports specifically on the translational
    motion of water molecules in the vicinity
    of a specially placed \gls{sp}
    -- a small molecule or moeity
    containing a stable unpaired electron
    spin (typically a nitroxide).
Specifically,
    sufficiently fast translation of the
    water relative to the \gls{sp} excites
    a transition that,
    combined with saturation of the \gls{sp}
    \gls{esr} transition,
    leads to \gls{nmr}
    signal inversion and enhancement.
\gls{odnp} reports the rate of this
    transition as a
    cross-relaxivity $k_\sigma$
    (with units of cross-relaxation rate per \gls{sp} concentration)
    that increases/decreases
    as the translational mobility of water
    within 1-1.5~nm of the \gls{sp}
    increases/decreases~\cite{Franck2015DNA,Barnes2017,FranckMethEnz2018}.

While previous \gls{odnp} studies have reported slow
    water motion at buried sites where relatively few
    water molecules can access the
    \gls{sp}~\cite{FranckMethEnz2018,Armstrong_apomb,Cheng_AnnRev2013},
    none have reported very low
    translational mobility near \glspl{sp} that were
    exposed to significant populations of water
    molecules.
Furthermore,
    one can imagine
    a strategy of functionalizing small spin
    probes designed to associate with
    particular environments, and then
    carefully ascertaining their environment
    with companion measurements.
Such a strategy has been relatively underutilized,
    with many \gls{odnp} measurements relying
    on either covalent attachment to
    macromolecules or intrinsically present
    radicals.
Past \gls{odnp} studies have investigated
    water dynamics in porous systems~\cite{Uberruck2018,Moon2022EviEntCon},
    and have applied \gls{odnp}
    to \gls{rm}-encapsulated proteins with a
    focus of using the combination of \gls{odnp}
    and \glspl{rm} as a vehicle for signal
    enhancement of the
    protein \gls{nmr}~\cite{Valentine2014}.
However, to the authors' knowledge,
    no systematic \gls{odnp} study has investigated
    water mobility as a function of confinement lengthscale.
The experimental results here dramatically underscore
    the reality that translational dynamics
    is a process relying on the collective
    rearrangement of relatively large masses
    of water.
%This does provide evidence for the idea that the
%translational motion of water is a collective
%process that requires several water molecules in
%order for it to take place % citations from talk.

\subsection*{\gls{sp} Localization by \gls{pre}-\gls{rosy}}
\label{sec:ODNPRMSpinProbes}
\figPREpieces
Basic chemical intuition indicates
    that the functional groups on small
    \gls{sp} molecules
    should guide the nitroxide into different
    microenvironments within a heterogeneous
    system:
    \eg, the organic \vs surfactant \vs water
    regions of the \glspl{rm} studied here.
However, validating the location of the \glspl{sp} can prove difficult.
For example,
    previous studies reported that
    zwitterionic \glspl{sp} have a preference for localization
    inside the internal water
    pool of \glspl{rm}~\cite{Haering1988}.
However, this result relied entirely on
    inferences from the \gls{esr} hyperfine coupling
    constants,
    rather than reporting directly on the
    approach of the nitroxide to specific
    chemical groups.

Here, we introduce \gls{pre}-\gls{rosy}
    for
    exploring the
    localization preferences of
    small \glspl{sp}.
\Gls{pre} is a well-established \gls{nmr}
    technique~\cite{Clore2009} in
    which the presence of \glspl{sp} causes
    nearby nuclei to relax more rapidly than they
    would in the absence of the \glspl{sp}~\cite{Clore2009}.
\gls{rosy} spectroscopy can illustrate the
    distribution of $T_1$ (\ie longitudinal)
    \gls{nmr} relaxation rates
    ($R_1=1/T_1$)
    both with and without the \gls{pre}
    arising from the incorporated nitroxide \gls{sp}.
The \gls{pre}-\gls{rosy} experiment
    (\cref{fig:PREpieces}),
    resolves different compounds according to their
    chemical shift (along the $x$-axis),
    while the displacement along the $y$-axis indicates
    how frequently the compound
    in the mixed-phase system
    comes close to the \gls{sp}
    (\cref{eq:DipolarRelax}).

\cref{fig:PREpieces} investigates
    a \gls{rm} of $w_{0} = 8$,
    with \gls{aot} surfactant
    and
    CCl$_{4}$ dispersant.
(CCl$_4$ is both \gls{nmr} silent
    and well-documented~\cite{harphamWater2004,tanDynamics2005,DAngelo1996}.)
It shows \gls{rm} samples containing water
    with dissolved TEMPOL (blue)
    \vs TEMPO-SO$_{4}$ (red),
    \vs the control
    (water with no \gls{sp}, purple).
It presents significant
    evidence that
    (anionic) TEMPO-SO$_{4}$
    localizes better to the center of the
    \gls{rm}'s internal water pool
    \vs
    (hydrophilic and neutral) TEMPOL.
First, the \gls{pre}
    (the displacement along the $y$-axis away from the
    purple contour)
    is significant for both hydroxytempo (blue)
    and tempo-sulfate (red),
    indicating that both \glspl{sp} come into close contact
    with the water;
however,
    the \gls{pre}
    of the water pool
    by TEMPO-SO$_{4}$
    is significantly greater.
Second, the \gls{tms} peak,
    which serves as a proxy for the
    dispersant (which is NMR silent),
    experiences a meager \gls{pre}
    effect from both \glspl{sp},
    indicating both tend to stay away from
    the dispersant;
however, the \gls{pre}
    is smaller for TEMPO-SO$_{4}$,
    indicating that it approaches the
    dispersant (where the \gls{tms} resides) less frequently than
    TEMPOL.
Therefore, for studying the water dynamics of the internal water pools,
    TEMPO-SO$_{4}$ was identified as the \gls{sp}
    better localized to the center of the water pool.
% ☐ TODO (JF): maybe say this is in contrast to some
% \gls{sp} - stearates?
% ☐ TODO (JF): something up with subfigure labels here check out later

% Want to include work that shows \gls{rm} is not perturbed by \gls{sp}
The SI further discusses the
    \gls{pre}-\gls{rosy} (\cref{sec:PREinSI}),
    additional deuterium \gls{rosy}
    (\cref{sec:ODNPRMSpinProbes2H}),
    and \gls{esr}~\cite{Haering1988,Hauser1989}
    (\cref{sec:ODNPRM_esr}) measurements
    that confirm these results.
\subsection*{\gls{odnp} Measurements of Translational Diffusivity}
\label{sec:ODNPRM_odnp}

\figkSigma

After proper background subtraction of the \gls{aot}
    signal (\cref{fig:CompareEnh}),
    \cref{fig:Ep}
    shows the \gls{odnp} enhancements for the four different $w_{0}$
    \glspl{rm};
    where enhancements, $E(p)$, are the ratio of \gls{nmr} signal
    with resonant microwaves, and therefore \gls{odnp},
    \vs without.
The lowest water loading (blue, $w_{0} = 1$) shows
    only enough \gls{odnp} to slightly
    reduce the \gls{nmr} signal,
    and not even enough to invert it.
This is consistent with a
water matrix whose translational motion
is dramatically restricted.
%Notably, however,
%    (upon zooming in, not shown)
%    the signal amplitude still follows a
%    clearly asymptotic power dependence indicating
%    clear saturation of the \gls{esr} transition.
The increasingly greater enhancements
    at increasing $w_{0}$
    indicate that the larger size of the
    enclosed water pools enables faster
    translational diffusion of the water
    molecules.
Overall, it is surprising that the \gls{odnp} turns
    on and almost completely off as the size
    of the water pool is scaled.

\tabKsig

In order to directly relate to translational
    diffusion,
    following established
    analyses~\cite{FranckMethEnz2018}
    (\cref{eq:polarization_subst_fundamental0}),
    the enhancements are
    subtracted from 1 and
    divided by the
    \gls{nmr} relaxation times,
    \gls{sp} concentration,
    and a standard set of physical constants
    to give \cref{fig:ksigma}.
These datapoints,
    in turn, are fit to an asymptotic microwave power
    dependence (to model the saturation of
    the \gls{esr} transition, \cref{eq:esrSat})
    to yield
    the cross-relaxivity or $k_{\sigma}$
    of each sample:
    \cref{tab:table_odnp}.
The translational retardation
    factor,
    $k_{\sigma,bulk}/k_{\sigma}$,
    in the second column
    indicates the slowness/retardation of the
    translational water diffusion relative to
    bulk~\cite{FranckMethEnz2018,Barnes2017,FranckPNMRS}
    and spans two orders of magnitude
    as the size of the \gls{rm} is altered.

The translational
    diffusion of water inside the \gls{rm}
    reduces to essentially
    non-existent levels
    in the smallest water pools
    ($k_\sigma$ $320\times$ smaller than in bulk,
    indicating correspondingly slower water motion).
This supports a view
    that translational motion of water is a
    collective process that requires hundreds of water
    molecules in order to achieve standard
    rates of diffusion~\cite{moilanenWater2009,Cerveny2016}.
At very small water loadings, the reduced
    number of water molecules do not provide
    a sufficiently large number of hydrogen-bonding
    configurations (microstates) that can be traversed
    in order to activate translational
    motion.
As the water pool grows in size, more
    water molecules are present to enable a larger
    number of hydrogen bonding configurations,
    thus enabling faster
    translational diffusion.

At the highest
    $w_{0}$ studied here,
    the water still moves less rapidly than in the
    bulk.
Notably, these measurements can be compared
    against water in the hydration layers of macromolecules
    dissolved in bulk water solution~\cite{Franck_crowding,FranckMethEnz2018},
    and a value of $k_{\sigma,bulk}/k_{\sigma}\approx 5$
    indicates that the water moves at rates
    comparable
    to the hydration water coating the surface of a lipid
    bilayer~\cite{Franck_crowding}.
At the highest water loading here,
    $w_0=10$, the diameter of the water pool should be
    about 3.5~nm~\cite{Pal2011FluCorSpe}.
(The value of $k_{low}$, \cref{eq:kLow}, for $w_0=10$ also
    matches $k_{low}$ at the surface of a
    lipid bilayer~\cite{Franck_crowding,FranckMethEnz2018}.)
However, unlike the lipid bilayer measurements
    that employ a \gls{sp} covalently bonded to the surface of the lipid
    bilayer~\cite{Franck_crowding},
    the spin label in the current measurements
    spends most of its time near the middle of the enclosed water pool.
Specifically, note that \cref{fig:PREpieces}
    shows a significantly greater \gls{pre}
    of the water \vs surfactant,
    thus indicating the \gls{sp} spends little time near the surfactant.

Furthermore,
    $k_{low}$
    (determined from the
        self-relaxivity, with the
        fast-motional contribution subtracted
        out: \cref{eq:kLow}),
    samples the dynamics
    of water in frequency space at a frequency
    that is $660\times$ smaller
    than that of $k_\sigma$~\cite{FranckMethEnz2018}.
($\sim 660$ is the ratio between the \gls{esr} and
    \gls{nmr} frequencies.)
Thus, both fast and slow
    motions
    contribute to $k_{low}$,
    with the
    slow motions
    likely including (or dominated by)
    rotation of the entire \gls{rm}.
The value of $k_{low}$ serves as a control to
    see whether or not
    the water molecules are still experiencing dipolar
    coupling with the \gls{sp}.
In all cases, the values of $k_{low}$ in \cref{tab:table_odnp}
    demonstrate good interaction of the water molecules
    with the \gls{sp}.
Thus,
    $k_\sigma$ falls off at low $w_0$ not
    because of any lack of magnetic dipolar
    coupling between the \gls{sp} and water,
    but because the
    translational motions of the water are simply not faster than the
    hundreds-of-picosecond timescales needed
    to achieve significant excitation of the
    cross-relaxation
    ($k_\sigma$) transition~\cite{Barnes2017,FranckMethEnz2018,Franck2014GroES,Franck2015DNA}.

\figRadiusPlot

Overall, these measurements provide greater insight into
    water that occurs naturally in
    confinement,
    ranging from pockets on the surface of
    proteins,
    to inside cellular vesicles (\eg lysosomes
    and endosomes),
    to the pores of synthetic mesoporous systems,
    and could pave the way for improved
    understanding of the function of these systems.
\cref{fig:RadPlot} emphasizes
    the near-total shutdown of translational diffusion for
    water pools of diameter $\sim 1.75\;\text{nm}$.
As previously noted,
    for the small water pools present at low water
    loadings,
    the dearth of different hydrogen bonding
    rearrangements possible leads to a slowdown in
    the translational motion of the water molecules.
However, the lengthscale at which this effect sets in
    is surprising:
    a \gls{rm} of $w_0=4.5-5$ encapsulates hundreds of water
    molecules~\cite{Ueda1988MeaAggNum,Beaton2023RapScrCor}.

% the following is correct, it's 30.76 nm when I zoom in
% derivative of slope is 0.0383 kσ/kσbulk vs w₀
Linear extrapolation (\cref{fig:RadPlot}) to
    $k_\sigma/k_{\sigma,bulk}=1$
    indicates that
    a water loading of
    $w_0=31$ would be required
    before the water in the center of the pool diffused
    at rates approaching those seen in bulk
    solution.
Again, the relatively large size of the water pool
    required proves suprising,
    as \glspl{rm} of this lengthscale
    contains
    an aqueous core
    $\sim 11\;\text{nm}$ in diameter,
    containing tens
    of thousands of water molecules.

It is worth noting that these experiments were initially designed
    with the intention of determining the
    thickness of the hydration layer from the
    perspective of \gls{odnp} by looking for
    a moderate change in $k_\sigma$ with the
    size of the reverse micelle.
However, they ended up
    highlighting how
    dramatically the translational mobility
    depends on the level
    of confinement.
As such, this study also opens up the pathway to
    future measurements where \glspl{rm} serve
    to relate measurements of dynamics to a lengthscale
    of confinement,
    in particular for \gls{odnp}
    measurements run at lower resonance frequencies
    that can ostensibly detect the dynamics of the
    slower water inside severely confined
    (\textit{i.e.} small) \glspl{rm}.
It also
    clearly points to future measurements where
    \gls{pre}-\gls{rosy} can verify
    the location of a spin probe inside
    a small but water-filled pocket
    of a protein or other macromolecule,
    while \gls{odnp} can elegantly
    analyze the translational diffusivity
    within that pocket.

\appendix
\ifarxiv%(
\putbib
\end{bibunit}
% https://tex.stackexchange.com/questions/168169/options-for-supplementary-materials-in-preprint-version-revtex-arxiv
%%%%%%%%%% Merge with supplemental materials %%%%%%%%%%
\pagebreak
\onecolumngrid
\begin{center}
\makeatletter
\textbf{\large Supplemental Materials for:\\
\@title}
\makeatother
\end{center}
\twocolumngrid
%%%%%%%%%% Merge with supplemental materials %%%%%%%%%%
%%%%%%%%%% Prefix a "S" to all equations, figures, tables and reset the counter %%%%%%%%%%
\begin{bibunit}
\setcounter{equation}{0}
\setcounter{figure}{0}
\setcounter{table}{0}
\setcounter{page}{1}
\makeatletter
\renewcommand{\theequation}{S\arabic{equation}}
\renewcommand{\thefigure}{S\arabic{figure}}
\renewcommand{\thesection}{S\arabic{section}}
%\renewcommand{\bibnumfmt}[1]{[S#1]}
%\renewcommand{\citenumfont}[1]{S#1}
%%%%%%%%%% Prefix a "S" to all equations, figures, tables and reset the counter %%%%%%%%%%
\glsresetall
\let\maincref\cref
% {{{ defs used by SI
\newcommand\Csl{\texorpdfstring{\ensuremath{C_{SL}}\xspace}{[SL] }}
% {{{ ubpair
\newlength{\mylength}
\newlength{\mysecondlength}
\newcommand{\ubpair}[2]{%
\newcommand*{\ubpairtext}{\mbox{\begin{footnotesize} #2\end{footnotesize}}}
%\renewcommand{\ubpairtext}{\begin{center}\begin{footnotesize}#2\end{footnotesize}\end{center}}
%\settoheight{\myheightforsubtract}{\ubpairtext}
\settowidth{\mylength}{\ubpairtext}%
\settowidth{\mysecondlength}{\ensuremath{\displaystyle #1}}%
%\mbox{lengths \the\mylength \the\mysecondlength}
\setlength{\mylength}{\maxof{\mylength-\mysecondlength}{0pt}/2}% only shift if the box is greater than the math
%\mbox{half of max is \the\mylength}
\hspace{-\mylength}%
\ensuremath{{\color{dbluecolor}\underbrace{\normalcolor #1}_{\ubpairtext}}}%
\hspace{-\mylength}%
}
\newcommand{\obpair}[2]{%
\newcommand*{\obpairtext}{\mbox{\begin{footnotesize} #2\end{footnotesize}}}
\settowidth{\mylength}{\obpairtext}%
\settowidth{\mysecondlength}{\ensuremath{\displaystyle #1}}%
%\mbox{lengths \the\mylength \the\mysecondlength}
\setlength{\mylength}{\maxof{\mylength-\mysecondlength}{0pt}/2}% only shift if the box is greater than the math
%\mbox{half of max is \the\mylength}
\hspace{-\mylength}%
\ensuremath{{\color{dbluecolor}\overbrace{\normalcolor #1}^{\obpairtext}}}%
\hspace{-\mylength}%
}
\newcommand{\Cmacro}{\ensuremath{C_{macro}}\xspace}%
\newcommand{\kHH}{\ensuremath{k_{HH}}\xspace}%
\newcommand{\krhop}{\ensuremath{k_{\rho}(p)}\xspace}%
\newcommand{\Tw}{\ensuremath{T_{1,w}}\xspace}%
\newcommand{\DeltaTw}{\ensuremath{\Delta T_{1,w}}\xspace}%
% }}}
% }}}
\textit{Please note that all
    figure/table/equation
references that are} not \textit{prefixed
with ``SI'' refer to the main text.}
\section{Theory and Data Analysis}
\label{sec:ODNPtheory2}
A spin cross-relaxation ($k_\sigma$)
    transition is always active when an electron
    spin and water are present.
This transition is excited by molecular motion
    that is sufficiently fast to cause high
    frequency fluctuations of the magnetic
    dipole-dipole coupling between the electron
    and proton (hydrogen nuclei of the water)
    spins.
By saturating (strongly exciting)
    the \gls{esr} transition to push the spin states
    away from thermal equilibrium,
    \gls{odnp} makes this cross-relaxation visible,
    specifically resulting in the enhancement and
    inversion of the nuclear (proton)
    spin states.
By normalizing the enhancements against the
    \gls{nmr} relaxation and driving the \gls{esr} transition
    to saturation,
    we can determine the transition rate of this
    cross-relaxation and relate it to the translational diffusion
    near the \gls{sp}~\cite{Franck2015DNA,Barnes2017,FranckMethEnz2018}.

Thus, cross-relaxivity,
    $k_{\sigma}$ (units $\text{s}^{-1}\text{M}^{-1}$, rate per \gls{sp} concentration), measures
    fluctuations in local magnetic fields
    generated by
    the translational motion
    of water.
\gls{odnp} also offers a measurement of the low-frequency relaxivity,
    $k_{low}$, that senses motion on a slower timescale, such
    as proton exchange
    or, here, tumbling of the reverse micelle~\cite{Franck2015DNA,FranckMethEnz2018}.
The quantification of both $k_{\sigma}$ and $k_{low}$ relies on measurements of the
the integrated signal intensity $I(p)$
as a function of microwave power, $p$.
The $I(p)$ are typically normalized
by the signal in the absence of microwave
power $I(0)$ to give the
\gls{odnp} signal enhancements
$E(p)=I(p)/I(0)$.
The $E(p)$ relate to the cross-relaxivity $k_\sigma$ discussed in the main text by way of~\cite{FranckMethEnz2018}:
\begin{equation}
    1-E(p)
    =
    \ubpair{T_1(p)}{\scriptsize\textit{1/rate of thermal relaxation}}
    \obpair{k_\sigma s(p) \Csl
    \left| \frac{\omega_e}{\omega_H} \right|}{\scriptsize\textit{rate of hyperpolarization}}
    ,
    \label{eq:polarization_subst_fundamental0}
\end{equation}
As indicated by the standard protocol~\cite{FranckMethEnz2018},
inversion-recovery measurements of the \gls{nmr} relaxation times
$T_1(p)$ at 3-5 select microwave powers, $p$,
are interpolated,
allowing division
of $1-E(p)$ by
$\Csl \left| \omega_e/\omega_H \right| T_1(p)$,
where $\left| \omega_e/\omega_H \right|$ is a constant ($659.3$),
and \Csl gives the number of moles of \gls{sp}
per liter of aqueous solution
(here $\Csl=70\;\text{mM}$ for all samples).
This division accomplishes the transformation
from \cref{fig:Ep} to \cref{fig:ksigma}.
\Cref{fig:ksigma} displays
    $k_\sigma s(p)$;
    \ie, it depends on the
    saturation of the \gls{esr} transition,
    $s(p)$,
    in addition to the desired
    cross-relaxivity.
Therefore, as noted in the text,
    $s(p)$ is fit to:
    \begin{equation}
        s(p) = \frac{s_{max}}{p+p_{1/2}}
        \label{eq:esrSat}
    \end{equation}
    where $p_{1/2}$ is a fitting parameter
    that depends on the characteristics of
    the hardware,
    while $s_{max}$ indicates the maximal electron
    spin saturation.
Here, it is
    reasonable~\cite{Armstrong_apomb,FranckPNMRS,Robinson1994}
    to assume that the slow
    tumbling seen in \cref{fig:rm_esr}
    drives $s_{max}$ to 1 by way of a
    $^{14}N$ relaxation mechanism that mixes
    the saturation of the 3 hyperfine lines.

The value of $k_{low}$ can be determined by
    subtracting the fast-motional $k_{\sigma}$
    from the
    self-relaxivity~\cite{FranckMethEnz2018},
    $k_{\rho}$:
\begin{equation}
    k_{low} = \frac{5}{3} k_{\rho} -
    \frac{7}{3} k_{\sigma}
    \label{eq:kLow}
\end{equation}
where $k_{\rho}$ is defined as 
\begin{equation}
    k_{\rho} = \frac{T_1^{-1}(0) - T_{1,0}^{-1}(0)}{C_{SL}}
    \label{eq:kRho}
\end{equation}
where $T_1(0)$ is the $T_1$ relaxation time of the sample in
    consideration with a \gls{sp} concentration of $C_{SL}$ at no microwave power
    ($p=0$) and $T_{1,0}(0)$ is the $T_1$ relaxation time of this same sample
    without any \gls{sp} present (also at $p=0$).
The unitless parameters $\big( \frac{k_{\sigma}}{k_{\sigma,bulk}}\big)^{-1}$
and $\big(\frac{k_{low}}{k_{low,bulk}} \big)$ enable facile comparison to
existing \gls{odnp} measurements in the literature, as described
in~\cite{Barnes2017,FranckMethEnz2018}.
The bulk values of $k_{\sigma} = 95.4\;\relxvtunits$
and $k_{low} = 366\;\relxvtunits$ are used to complete this
analysis~\cite{Franck2014GroES}.

High-field (400~MHz) $^1\text{H}$
    \gls{pre}-\gls{rosy} measurements
    identify the location of the
    small-molecule \glspl{sp} in this study.
\Gls{pre} is a well-established \gls{nmr} technique in
    which the presence of nearby \glspl{sp} causes
    nearby nuclei to relax more rapidly than they
    would in the absence of the \gls{sp}~\cite{Clore2009}.
In this way, \gls{pre} can be used to obtain distance
    information in, \eg biological systems,
    where a covalently linked \gls{sp} can be used
    to report on nuclei near that \gls{sp}.
(Or where small molecule \gls{sp} in solution
    helps to identify surface-exposed sites.)

% \gls{pre} paragraph
As with \gls{odnp}, \gls{pre} is brought about by
dipolar interaction between an electron spin and a
nuclear spin.
Typically, for a \gls{sp} with $S=1/2$, like a
nitroxide, this takes the form~\cite{Clore2009}: 
\begin{equation}
    R_1-R_{1,0}=
    \frac{1}{T_{1}}
    -
    \frac{1}{T_{1,0}} =
    \frac{3}{10}
    \left( \frac{\mu_0}{4\pi} \right)^2
        \frac{
        {\hbar}^{2}{\gamma_{I}}^{2}{\gamma_{S}}^{2}
    }{{r}^{6}}
    \frac{\tau_{c}}{1+(\omega \tau_c)^2}
    \label{eq:DipolarRelax}
\end{equation}
where
    $T_1$ and $T_{1,0}$ are the \gls{nmr}
    relaxation times (and $R_1$ and
    $R_{1,0}$ the corresponding rates)
    with and without the
    spin label (here at high field and resonance frequency),
    $\gamma_{I}$ and $\gamma_{S}$ are the
    gyromagnetic ratios of the nuclear spin and
    electron spin, respectively, $r$ is the distance
    between the spins,
    $\omega$ is the resonance frequency
    (here $2\pi \times 400\;\text{MHz}$)
    and $\tau_{c}$ the correlation
    time of the relative motion of the spin
    probe and molecule of interest,
    under the assumption that the electron spin
    relaxes too quickly to be affected by motion on
    this
    timescale~\cite{bloembergenRelaxation1948,solomonRelaxation1955}.
Note that, for very fast motions,
    $\omega\tau_c$ will approach zero.

Of note in~\cref{eq:DipolarRelax} is the
    squared dependence on $\gamma_{S}$,
    making the relaxation arising from
    coupling to electron spins
    many orders of magnitude faster than that
    arising from other nuclear spins.
As such, dipolar relaxation in the presence of
    unpaired electrons tends to dominate all
    other relaxation mechanisms in proton
    \gls{nmr},
    where its contribution to
    $R_1=1/T_1$ is very large. 
Also of note,
    the $\gamma_I^2$ dependence of \cref{eq:DipolarRelax}
    suppresses the \gls{pre} effect for low-$\gamma$ nuclei.
Since $\gamma_{I}$ is ~6.5$\times$ smaller for
    $^{2}H$ than it is for $^{1}H$,
    and, furthermore,
    for $^2H$,
    a quadrupolar mechanism drives
    a relatively large contribution
    to the $1/T_1$ rate,
    the \gls{pre} contributes negligibly in
    deuterium
    \gls{nmr}~\cite{mantschDeuterium1977}.

In this study,  because both the spin probe and
    the molecules experiencing \gls{pre} are
    relatvely small in size
    (compared to \textit{e.g.} proteins that
    are also studied by \gls{pre}),
    it is important to note that
    \cref{eq:DipolarRelax} scales with
    $1/r^6$,
    so close-contact interactions will tend dominate
    the \gls{pre} interaction.
The \gls{pre} in this context can be said
    to be a measure of how often a given
    molecule or group comes close to the
    \gls{sp}.
\subsection{Size of \glspl{rm}}
Various points in the manuscript reference the
    number of waters present inside the \gls{rm}.
While there are various schemes for estimating
    this number,
    as noted in~\cite{Beaton2023RapScrCor},
    we employ a meta-analysis of several
    sources~\cite{Ueda1988MeaAggNum,Eskici2016SizAOTRev,Amararene2000AdiComAOT,Eicke1976ForWatOil}
    to arrive at the equation
    for the number of AOT molecules (aggregation number $\bar{n}$) per
    \gls{rm}:
\begin{equation}
    \begin{array}{rl}
        \displaystyle
        \exp\left( k \bar{n} \right)
        =&
        \exp\left( k a \right)
        +
        \exp\left( k (m w_0 + b) \right)\\
        \displaystyle
        &
        +
        \exp\left( k (l w^2_0 + c) \right)
    \end{array}
    \label{eq:numWaters}
\end{equation}
From $\bar{n}$, one can in turn determine
    the number of water molecules, $\bar{n}w_0$,
    at a given water loading.
In \cref{eq:numWaters}, the value of $a=15.1$ gives number of AOT molecules
in the constant-$\bar{n}$ regime,
the value of $m=7.15$ gives the slope of the linear region
(number of AOT molecules added per change in $w_0$),
$b=0.259$ the intercept of the linear regime,
$l=0.673$ the curvature of the quadratic
regime,
$c=-37.1$ the intercept of the quadratic
equation,
and the value of $k=0.174$ gives the curvature between
the constant, linear, and quadratic regimes.
\section{Experimental}
\label{sec:ODNPRMexp}
\subsection{Sample preparation}
Two \glspl{sp} were studied in this work: TEMPOL
    (4-hydroxy-2,2,6,6-tetramethylpiperidin-1-oxyl),
    and the ionic TEMPO-SO$_{4}$ (4-sulfate-2,2,6,6-tetramethylpiperidin-1-oxyl
    potasium salt), as shown
    on the left side of \cref{fig:PREpieces}.
The nitroxide moiety is located on a
    neutral, polar molecule
    in TEMPOL
    and on a negatively charged ion in TEMPO-SO$_{4}$.

TEMPO-SO$_{4}$ was synthesized using a
    previously described
    procedure~\cite{winsbergAqueous2017}.
The lowest and highest $w_{0}$ \glspl{rm} were prepared first, and intermediate
    $w_{0}$ samples were prepared by mixing
    appropriate aliquots from each in order to
    keep the concentration of \gls{aot} constant across
    all sample preparations.
    The constant \gls{aot} concentration (as opposed to a constant water concentration)
    in all samples better suits the ability to
    background subtract contributions to the \gls{nmr} signal.

The lowest water loading sample ($w_{0} = 1$) was prepared by
    adding \gls{aot} (4.37~g, 0.00983~mol) to
    6.765~mL (10.78~g)
    CCl$_{4}$. The clear solution was vortexed until
    all \gls{aot} was dissolved, after which
    179~μL of 70 mM TEMPO-SO$_{4}$ was added to
    the solution and vortexed 30 seconds
    3$\times$.
The highest water loading sample ($w_{0} = 10$) was prepared by
    adding \gls{aot} (4.44 g, 0.00999 mol) to 6.765 mL
    CCl$_{4}$ and dissolving completely, after which
    1800~μL of 70 mM TEMPO-SO$_{4}$ was added to
    the solution and vortexed for 30 seconds
    3$\times$.
After vortexing, each sample was capped, wrapped in parafilm, placed in a
    desiccator, and allowed to sit at room temperature
    for several hours.
    The final solutions had a slight red color
    (arising from the nitroxide),
    most noticeable in the $w_{0} = 10$ sample.
These preparations correspond to mass percentages
    of 63\%, 29\%, and 11\% for the $w_{0} = 10$ sample
    and 70\%, 29\%, and 1\% for the $w_{0} = 1$ sample
    for CCl$_{4}$, \gls{aot}, and water, respectively. 
Given that these samples were prepared in CCl$_{4}$, the highest $w_{0}$
    attainable is around 10 \cite{fennWater2011}.

To prepare the $w_{0} = 5$ sample, 2070~μL of
    each the $w_{0} = 1$ and $w_{0} = 10$ samples were
    combined to give a final solution volume of 4140~μL;
to prepare the $w_{0} = 7$ sample, 1320~μL of
    each the $w_{0} = 5$ and $10$ samples were
    combined to give a final solution volume of 2640~μL;
    \etc
These solutions were vortexed for 30 seconds
    3$\times$ and allowed to become fully
    transparent before mixing subsequent aliquots
    or preparing \gls{nmr} samples.
After attaining full transparency, each sample
    exhibited a slight red color as with the initial
    $w_{0} = 10$ solution.

To prepare the parallel D$_{2}$O samples, the same
    preparation was followed with equivalent amounts
    of a 70 mM TEMPO-SO$_{4}$ D$_{2}$O solution used
    in place of the H$_{2}$O solution.
The parallel D$_{2}$O samples provide
    background signal to account for any enhancement
    of the \gls{aot} protons that contribute to the detected
    signal, as in the work by Valentine and
    coworkers~\cite{Valentine2014},
    and as shown in \cref{fig:CompareEnh}.
\subsection{\gls{odnp} measurements}
\figCompareEnh
\gls{odnp} measurements were performed on an
    system comprising
    a home-built \gls{nmr} probe that inserts into a
    Bruker Super High Sensitivity
    Probehead X-Band resonator (ER 4122
    SHQE, probe as in \cite{FranckPNMRS}),
    of a Bruker E500 cw EPR,
    a home-built receiver box
    (with home-built duplexer,
    Minicircuits ZFL-500LN+,
    and various commercial filters),
    a SpinCore RadioProcessor-G transceiver
    board to digitize the signal,
    and a Bridge12 \gls{mps}.

The field concomitant with the lowest field
    hyperfine line (see \cref{fig:rm_esr})
    was used as the $B_{0}$ field for
    \gls{odnp} measurements.
The exact field position was determined before
    each \gls{odnp} experiment by performing a field
    sweep of the \gls{nmr} resonance with 3 W of
    incident microwave power, thus identifying the
    field position at which maximum enhancement
    occurs (approximately 3490 G for a microwave
    frequency of 9.8~GHz).

The python extension for
    controlling the SpinCore transceiver with
    a simple list-based pulse program format
    and automating the \gls{odnp} experiments
    is available at
    \url{https://github.com/jmfrancklab/spincore_apps}
    this relies in part on
    the library available at
    \url{https://github.com/jmfrancklab/inst_notebooks},
    which provides utilities for interfacing
    with the \gls{mps} (to control the microwave power)
    as well as the Bruker
    magnet (to change the field),
    and provides continuous logging of the
    microwave power incident on the sample.
All data is stored in the standard HDF5
    (Hierarchical Data Format)
    format.

Additional tools available in
    \url{https://github.com/jmfrancklab/proc_scripts}
    help to process signal following the
    procedure outlined
    in~\cite{Beaton2022ModVieCoh}.
All the home-built libraries mentioned here utilize
    the pySpecData
    (\url{https://github.com/jmfrancklab/pyspecdata})
    library,
    as it simplifies and clarifies all key
    tasks related to manipulating
    multi-dimensional spectroscopic data.
\subsection{\gls{pre}-\gls{rosy} measurements}
\gls{pre} measurements were acquired with a
    custom inversion-recovery sequence
    on an Avance III HD 400 MHz.
The variable delay list (for recovery delay
    $\tau$) had roughly
    logarithmically spaced points,
    to capture both faster and slower relaxing
    components.
Both to improve the signal-to-noise,
    and because the $J$-coupling multiplets of the surfactant
    appear different for different levels of $T_1$ relaxation,
    the spectral dimension is subjected to significant
    Gaussian apodization (0.05~ppm).
The $R_1$ distributions were implemented
    by following a 1D version of the
    \gls{ilt}
    outlined in
    \cite{Venkataramanan2002SolFreInt},
    in particular utilizing the compressed implementation of
    the BRD~\cite{butler_estimating_1981} method
    for choosing the regularization parameter.
The algorithm is available as part of the
    pySpecData library (\texttt{nnls} function).
For the \gls{pre} data,
    a basis of linearly spaced $R_1=1/T_1$
    values was chosen,
    with kernel $K(R_1,\tau)=\exp(-R_1\tau)$.
Samples for \gls{pre}-\gls{rosy} followed the same general scheme as noted
    above,
    except that trace amounts of \gls{tms} were added to the sample.
\subsection{Further notes on determination of $k_{low}$}
Note that to determine $k_{low}$,
    a non-\gls{odnp}-enhanced \gls{nmr} experiment must
    measure the relaxation rate in the absence of \gls{sp}.
As this measurement involves a prohibitively low level
    of \gls{snr}, it was only performed for the highest
    $w_0$ \gls{rm}, where water is expected to have the
    longest relaxation time.
Thus, \cref{tab:table_odnp} indicates the lower bound for
    $k_{low}$ in all cases.
\section{Further notes on \gls{pre}-\gls{rosy}}\label{sec:PREinSI}
In \cref{fig:PREpieces}, it is worth noting
    that the surfactant peaks
    experience slightly less
    \gls{pre} from the TEMPO-SO$_4$ \vs
    TEMPOL that becomes more apparent upon
    zooming in on the relevant peaks.
This is consistent with the general result
    that the TEMPO-SO$_4$ remains better
    localized to the middle of the water
    pool,
    away from both dispersant and surfactant.

It is also worth noting that at any instant,
    multiple different populations of \gls{rm}
    coexist in equilibrium:
    one of \glspl{rm} loaded with
    a single \gls{sp},
    another with no \gls{sp},
    a third with two \glspl{sp},
    \etc.
The exchange of water
    from one \gls{rm} to another should be
    relatively slow.
Each population of \glspl{rm} with different numbers of
    encapsulated \glspl{sp}
    provides a different micro-environment for the
    water,
    which will relax more rapidly inside \glspl{rm} with
    more \glspl{sp}.
The \gls{rosy} spectrum,
    however,
    shows only one peak in the $T_1$ distribution,
    indicating that the slow exchange of water
    between the different \gls{rm} microenvironments
    still occurs on a
    timescale that is faster than the
    $T_1$ of the (\gls{sp}-labeled)
    \gls{rm} -- \ie,
    faster than about 100~ms.
Therefore, the \gls{rosy} results also highlight that
    measurements of \gls{odnp}
    cross-relaxation ($k_\sigma$)
    will be averaged across all \glspl{rm}
    in the ensemble.

\section{$^2$H \gls{rosy} to check for structural
perturbation of Water}
\label{sec:ODNPRMSpinProbes2H}

As an important control,
    the deuterium relaxometry
    screening technique (\gls{adrosys}) described in~\cite{Beaton2023RapScrCor}
    was applied to study the TEMPO-SO$_{4}$ loaded
    \gls{rm} system,
    to check whether the \gls{sp} might be
    significantly perturbing the properties
    of the confined water.
In this way, the \gls{sp} acts
    like the inclusion molecules studied
    in~\cite{Beaton2023RapScrCor}.
This technique
    could offer insight into any changes to the
    hydrogen bonding strength or rotational mobility
    of water in this \gls{rm} system.
Importantly, unlike $^{1}$H nuclei, however,
    $^{2}$H nuclei do not
    experience a dramatic \gls{pre} effect in the
    presence of unpaired
    electrons~\cite{mantschDeuterium1977}
    (due to the $\gamma_I^2$ scaling in \cref{eq:DipolarRelax}).
As a consequence, the resulting relaxometry data
    should not reflect much of a change relative to the
    ``empty'' (water-only) \gls{rm} system.

\figCarbonTetSL

\cref{fig:CtetwithSL}
    presents data for TEMPO-SO$_{4}$ loaded \glspl{rm}
    with sample preparations ranging
    from $w_{0} = 1.0$ to 8.1.
The expected behavior is observed -- namely,
    that there is an increase in $^2$H $T_1$ and
    chemical shift as $w_{0}$ increases.
When compared to the data of~\cref{fig:justCtet}
    in~\cite{Beaton2023RapScrCor}, from 
    \glspl{rm} of similar sizes in the same dispersant
    (CCl$_{4}$) but without any \gls{sp}, there is
    general agreement.
One slight difference is
    that upon incorporation of the \gls{sp}
    (compare \cref{fig:justCtet} \vs \cref{fig:CtetwithSL}),
    the distribution for the $w_0=1$ broadens somewhat.
This is consistent with the data obtained in the
    presence of guest molecules present in high
    concentrations, such as the glucose and PEG-200
    data reported
    in~\cite{Beaton2023RapScrCor},
    and could indicate a change in the size
    distribution of the \glspl{rm} resulting
    from inclusion of the guest molecule.
The slight change in chemical shift seen for all water loadings
    (compare \cref{fig:justCtet} \vs \cref{fig:CtetwithSL}),
    is likely consistent with a paramagnetic
    shift upon incorporation of the \gls{sp}.
Overall, this data thus indicate that the \gls{sp} does not dramatically perturb the rotational
    motion or hydrogen bonding network of the internal water pool at the
    concentration (70~mM) used.

\subsection{Electron Spin Resonance}\label{sec:ODNPRM_esr}
\figRMESR
\tabRMESR
The \gls{rm} samples were also analyzed by cw \gls{esr}
    at 9.8~GHz in \cref{fig:rm_esr}.
In line with previous
    measurements~\cite{Haering1988,Hauser1989},
the \gls{esr} spectra show a restriction of the \gls{sp} mobility at the lowest
    water loadings ($w_0 = 1$, blue) and an incremental increase in the mobility
    as the $w_0$ increases, up to the highest $w_0$ in these measurements
    ($w_0 = 10$, red).
Specifically, the
    \gls{esr} spectra for nitroxide \glspl{sp}
    inside \glspl{rm} exhibit a characteristic
    decrease in
    intensity of the three hyperfine lines from
    lowest to highest field
    (left to right)
    as well as a general broadening of all
    lines
    due to the reduced rotational correlation
    time of the \gls{sp}.
(Note that \gls{esr} spectra are
    presented as a derivative of the
    absorption spectra,
    so that each positive-negative pair in
    the lineshape is referred to as a
    ``line.'')
The difference from nitroxides freely
    dissolved in either water or organic solvent
    at room temperature,
    which present three roughly equal lines,
    is consistent with the fact that at all
    $w_0$ studied here, \glspl{rm} were
    successfully formed and the
    \gls{sp} remained trapped inside the
    aqueous core of the \gls{rm}.
The equation~\cite{Caldararu1994StrPolCor}
\resizebox{\linewidth}{!}{%
    \(\displaystyle
        \tau_c = 
        \left( 
        6.51\times10^{-10}\;\text{s}
        \right)
        \left( 
        \frac{\Delta H_0}{\text{G}}
        \right)
        \left( 
            \sqrt{\frac{h(0)}{h(-1)}}
            +
            \sqrt{\frac{h(0)}{h(+1)}}
            -2
        \right)
    \)
}
approximates the rotational correlation times,
where s and G are units (seconds and
    Gauss);
    $h(+1)$,
    $h(0)$,
    and
    $h(-1)$
    are the peak-to-peak amplitudes of the
    \gls{esr} lines;
    and $\Delta H_0$ is the peak-to-peak
    linewidth of the central line.

The trend observed is opposite what one would expect from a
    Stoke-Debye-Einstein
    prediction of the rotational
    correlation time
    of the \gls{rm}
    (\ie, $\tau_R\propto R^{-3}$ with
    $R$ the \gls{rm} radius).
Rather, the decreasing correlation time with
    increasing \gls{rm}
    diameter is consistent with the notion that the water
    pool develops more bulk-like character with
    increasing $w_0$ and that the microviscosity
    decreases with increasing $w_0$.
Similar results have been observed previously
    for slightly different \glspl{sp}~\cite{Caldararu1994StrPolCor},
    and prove consistent with (though less
    direct than) the \gls{odnp} results
    presented here.
Notably, the spectrum for the
    highest $w_0$ studied here ($w_0 = 10$) still
    does not 
    exhibit hyperfine lines of equal intensity, as
    expected for a
    \gls{sp} free in solution, thus indicating that
    there is still some amount of rotational
    retardation.
However, for small $w_0$,
    as the water becomes very viscous and potentially
    almost completely immobilizes the \gls{sp}
    relative to the reverse micelle,
    the $\tau_c$ observed in the \gls{esr} spectrum
    will report almost exclusively on the rotational
    correlation time of the water.
\section{Other methods of extrapolation}
Extrapolating a quadratic fit of the $k_\sigma$
    \vs the water pool diameter
    indicates a water pool diameter of 6.7~nm is
    needed which is roughly $w_0 = 20$:
    although not quite as large,
    this gain confirms that a very large assembly of water
    molecules is needed to provide bulk-like
    translational dynamics.

\putbib
\end{bibunit}
\else%)(
%%%%%%%		References			%%%%%%% 
\setlength{\bibsep}{0.0cm}
\bibliographystyle{Wiley-chemistry}
\bibliography{references}
\clearpage
%%%%%%%		TOC Entry			%%%%%%% 
\section*{Entry for the Table of Contents}
%%%%%%%		 Option 2			%%%%%%%    
\noindent\rule{11cm}{2pt}
\begin{minipage}{11cm}
\includegraphics[width=11cm]{figures/water_zooming_rm.png}
\end{minipage}
\begin{minipage}{11cm}
%\large\textsf{Authors should provide a short Table of Contents graphical abstract and accompanying text (up to 450 characters including spaces). The graphical abstract should stimulate curiosity. Repetition or paraphrasing of the title and experimental details should be avoided.}
\large\textsf{%
Cartoon of a 360 ps dynamic snapshot of translational water diffusion in a confined environment,
with distance of water translation and all sizes approximately to scale.
In the least confined environment observed here,
Overhauser Dynamic Nuclear Polarization observes a diffusion about 5× slower than bulk water (left).
For confinements smaller than the correlation length of water,
it observes the near absence of translational diffusion (right).
}
\end{minipage}
\noindent\rule{11cm}{2pt}
%%%%%%%%%%%%%%%%%%%%%%%%%%%%%%%%%%%%%%%%%%%%%%%%%%%%%%%%%%
%%%%%%%%%%%%%%%%%%%%%%%%%%%%%%%%%%%%%%%%%%%%%%%%%%%%%%%%%%
%%%%%%%%%%%%%%%%%%%%%%%%%%%%%%%%%%%%%%%%%%%%%%%%%%%%%%%%%%
\fi%)
\end{document}